\documentclass[twocolumn,showpacs,preprintnumbers,prl]{revtex4}
\usepackage{graphicx}
\usepackage{dcolumn}
\usepackage{bm}
\usepackage{units}
\begin{document}
\draft
\title{Inhomogeneous High Harmonic Generation in Krypton Clusters}
\author{H. Ruf$^{1}$, C. Handschin$^{1}$, R. Cireasa$^{2}$, N. Thir\'e$^{2}$, A. Ferr\'e$^{1}$, S. Petit$^{1}$, D. Descamps$^{1}$, E. M\'evel$^{1}$, E. Constant$^{1}$, V. Blanchet$^{2}$, B. Fabre$^{1}$, Y. Mairesse$^{1}$}
\affiliation{$^{1}$Universit\'e de Bordeaux - CNRS - CEA, CELIA, UMR5107, F33405 Talence, France \\
$^{2}$ a Universit\'e de Toulouse, 118 route de Narbonne, F-31062 Toulouse, France\\ 
b CNRS, Laboratoire Collisions Agregats Reactivite, IRSAMC, F-31062 Toulouse, France}

\begin{abstract}
High order harmonic generation from clusters is a controversial topic: conflicting theories exist, with different explanations for similar experimental observations. From an experimental point of view, separating the contributions from monomers and clusters is challenging. By performing a spectrally and spatially resolved study in a controlled mixture of clusters and monomers, we are able to isolate a region of the spectrum where the emission purely originates from clusters. Surpringly, the emission from clusters is depolarized, which is the signature of statistical inhomogeneous emission from a low-density sources. The harmonic response to laser ellipticity shows that this generation is produced by a new recollisional mechanism, which opens the way to future theoretical studies. 
\end{abstract}

 \maketitle
\newpage

High harmonic generation (HHG) \cite{Lewenstein} refers to the interaction of high intensity laser light with matter, which leads to the emission of broadband coherent radiation in the extreme ultraviolet domain. HHG from clusters is considered a promising light source, showing higher emission frequencies \cite{Donnelly, Vozzi}. HHG is also a spectroscopic tool to extract structural and dynamical information on the emitting medium from the properties of the harmonic radiation (spectrum, phase and polarization state). This technique, which has been used to probe atoms \cite{Worner, Higuet} and small molecules \cite{Marangos}, relies on the basic mechanism of HHG described as a three step model \cite{Krause, Corkum, Lewenstein}: First, a bound electron escapes in the strong laser field through tunnel ionization. Second, the electron is driven away then accelerated back towards the parent ion. Finally the electron recombines radiatively with the parent ion. This recombination encodes the structure of the medium in the emitted light. The extension of this technique to the case of clusters would allow investigation of strong field processes in many-body systems, the role of multielectron effects and monitoring of cluster dynamics through the harmonic signal \cite{Strelkov, Chen}. 

The exact mechanism of HHG from clusters is still debated. Various extensions of the three step model have been proposed \cite{Moreno, Hu, Veniard, Zaretsky}. The dominant channel generally considered is ionization and recombination to the same atom (atom-to-itself). Since clusters are dense media, there is also a possibility of recombination to neighbouring ions \cite{Moreno, Veniard}. This atom-to-neighbour emission can produce incoherent radiation due to a lack of phase-locking between the two atomic wavefunctions \cite{Zaretsky}. Another contribution to harmonic emission may come from a wavefunction partially delocalized over the whole cluster, from which electrons tunnel out of and to which they recombine coherently (cluster-to-itself). In addition to these recollisional mechanisms, very different mechanisms could also co-exist as is the case in overdense plasmas \cite{Quere} or bulk crystals \cite{Ghimire}. From an experimental point of view a particular difficulty consists in disentangling the harmonics produced by different species (monomers and different size clusters) and possibly different mechanisms.

In this letter we present a detailed experimental study of HHG from clusters which aims at disentangling the harmonic signal from clusters and identifying the mechanisms at play. By studying the spectral and spatial profile of the harmonics as a function of pre-expansion gas temperature, we identify one region corresponding to pure emission from clusters. A simulation verifying the results from a polarization measurement confirm that only few emitters contribute to this region, which is consistent with the low density of large clusters in the generating medium. The fast decay of the harmonic signal with laser ellipticity indicates that these harmonics are produced by a recollision mechanism which strongly depends on the cluster size, as expected in a cluster-to-itself picture.

We generate high harmonics in a supersonic gas jet, employing a valve (Even-Lavie \cite{Luria}) pulsed at 1\,kHz with a $d=150\,\mu m$ diameter trumpet nozzle and a jet expansion half angle of $\alpha\sim 10^{\circ}$. The backing pressure was kept constant at $p_0=20\,$bar, while the pre-expansion nozzle temperature $T_0$ varied. Cluster formation can be described by the empirical Hagena scaling parameter $\Gamma^*=k\,\frac{(d/tan\alpha)^{0.85}}{T_0^{2.29}}\,p_0$ \cite{Smith}, with a condensation parameter $k$ equal to 2890 for krypton~\cite{Smith}. The average cluster size can be estimated by~\cite{Dorchies}: $\bar{N}=33\left(\Gamma^*/1000\right)^{2.35}$. In our experiment the cluster size was varied by changing the nozzle temperature between 350\,K ($\bar{N}\approx33000$) and 510\,K ($\bar{N}\approx4000$). The standard deviation $\delta N$, of the cluster size distribution is broad and leads to significant contribution of small cluster, which can be estimated by $\delta N / \bar{N} = 45\%$ \cite{Dorchies}. 

\begin{figure}
\begin{center}
\includegraphics[width=0.5\textwidth]{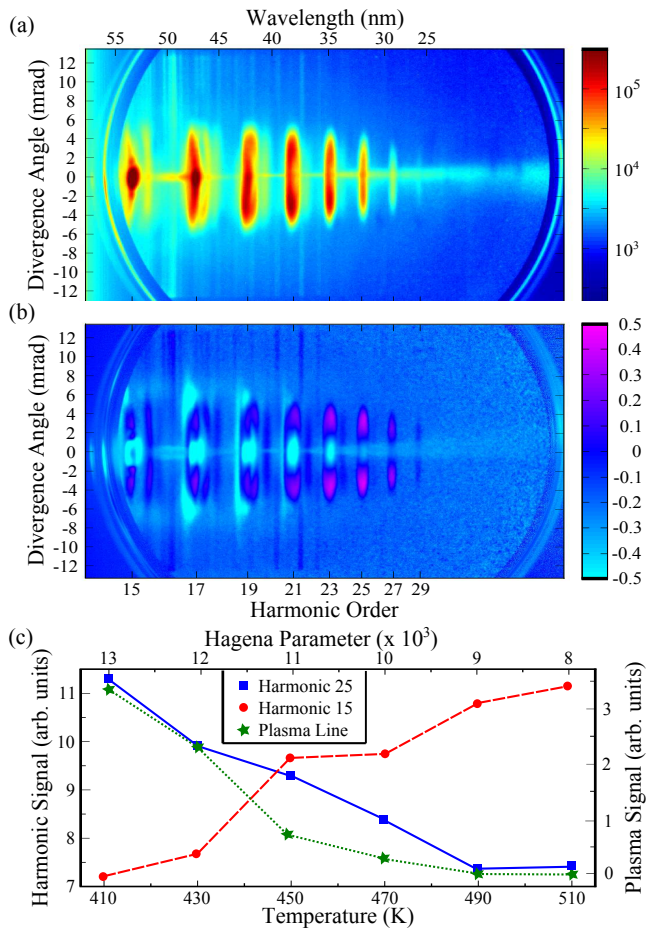}
\end{center}
\caption{(a) High harmonic spectrum at 350\,K. The vertical lines correspond to electronic transitions in krypton ions. (b) Normalized difference signal of a spectrum at 350\,K and a spectrum at 510\,K. The signal from the pink areas dominates at 350\,K.}
\label{fig:spec}
\end{figure}

We used the Aurore Ti:Sa-Laser system from CELIA which delivers 7 mJ, 35 fs, 800 nm pulses at 1 kHz. The laser was focused using a 37.5\,cm focal spherical mirror at 4.8\,mm from the nozzle exit, which ensures proper jet thermalization. The harmonic spectrum was dispersed using an abberation-corrected concave gold grating. The imaging detector consisted of dual microchannel plates coupled to a phosphor screen and a 12 bit CCD-camera. Each image was averaged over 15000 laser shots.  

Figure \ref{fig:spec}(a) shows a harmonic spectrum obtained at a laser intensity of 1$\times$10$^{14}$\,W/cm$^2$. The horizontal line around zero divergence angle is due to scattered light from lower harmonics. In addition to the harmonic spectrum, we observe isotropic radiation as vertical lines, corresponding to plasma lines~\cite{McPherson,Ditmire}, which originate from electronic transitions in excited krypton ions. Plasma lines at 38.7\,nm and lower wavelengths correspond to electronic transitions in at least five times ionized krypton ions~\cite{NIST}. A strong line observed at 43.4\,nm refers to an electronic transition in Kr$^{7+}$~\cite{Reader}. The laser intensity employed is not high enough to field ionize krypton atoms five (I$_p$=78.5\,eV) or seven times (I$_p$=125.8\,eV). In turn, impact ionization within clusters leading to inner shell ionzation \cite{Krainov} and leaving the freed electrons in a quasy bound cloud behind, can be responsible for the formation of such highly ionised Kr~\cite{Hu, Fennel}. 

Apart from plasma lines, the spectrum from Fig.~\ref{fig:spec}(a) does not show specific features that could be the signature of the presence of clusters in the medium. In order to identify the contribution of clusters we performed a differential measurement: we monitored the spatially and spectrally resolved evolution of the signal as the nozzle temperature increased from 350 K to 510 K, which decreases the cluster average size. Fig. \ref{fig:spec}(b) reveals clear regions with distinct behaviors: a low energy component (up to harmonic 25) which increases with temperature, and an off-axis component (dominant from harmonic 23 to 31) which presents a higher cutoff and decreases monotonously with temperature (Fig.~\ref{fig:spec}(b)) as the plasma line. In the following we will focus on the latter component, whose behavior is correlated to that of the plasma lines and thus to the presence of large clusters. 

\begin{figure}
\begin{center}
\includegraphics[width=0.5\textwidth]{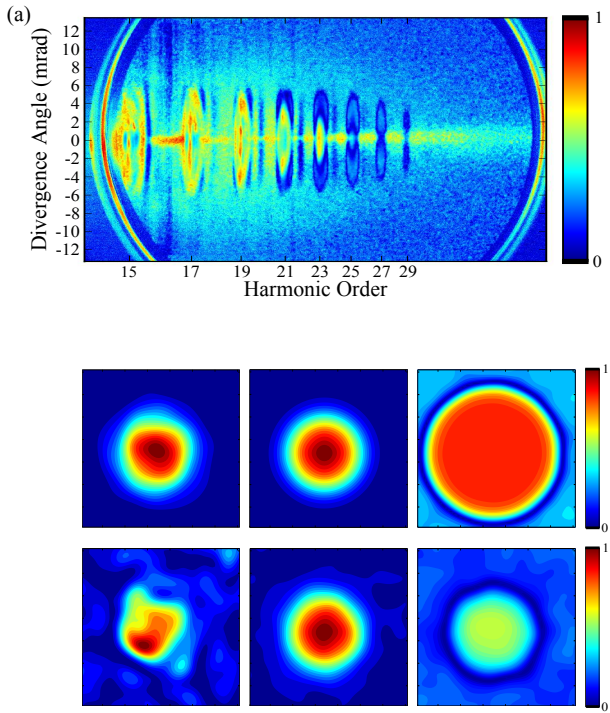}
\end{center}
\caption{(a) Polarimetry map depicting the oscillation amplitude of the the Malus' law for each pixel at 350\,K. (b) (b) Simulated harmonic spatial profile for a single laser shot (left), 1000 shots (middle), and degree of linear polarization for 1000 shots (right). The first row corresponds to HHG in atoms and the second row to HHG in large clusters.}
\label{fig:polar}
\end{figure}

In order to characterize the different mechanisms that may be at play in the harmonic emission, we performed a polarimetry measurement~\cite{Antoine}. We used an unprotected silver mirror in combination with a gold grating as a fixed polarizer and rotated the laser polarization with a zero-order half waveplate. The signal of each pixel of our detector is monitored as a function of the angle of the half waveplate resulting in a Malus' law with a $cos^2$ dependence. After subtracting the background, a Fourier transform is performed to extract the amplitude of the oscillatory component, which is then normalized to the sum of the Fourier transform. This procedure provides the amplitude of the oscillation and therewith the degree of linear polarization. Fig. \ref{fig:polar}(a) depicts the spatio-spectrally resolved degree of linear polarization obtained at 350\,K. As expected, the plasma lines are clearly unpolarized. More surprisingly, the same region we attributed to cluster emission in Fig. \ref{fig:spec}(b) shows a remarkably low degree of linear polarization. A low degree of linear polarization is very unexpected for high harmonic generation by a linear laser field \cite{Antoine}. Note that we systematically observed this low degree of linear polarization in different conditions and various species (CO$_2$, argon and mixtures of CO$_2$/krypton), but not in the absence of clusters using an effusive gas jet.

The polarization direction of the high harmonics is set by the recollision direction of the electrons and the electronic structure of the ground state to which these electrons recombine. Within the strong field approximation, in linear polarization the recollision direction is parallel to the laser field. In a centrosymmetric medium the harmonics are thus necessarily polarized along this direction. To go beyond the strong field approximation, we have performed Classical Trajectory Monte Carlo (CTMC)~\cite{Higuet} calculation of the recollision angles in atoms at intensities of 1.2 $\times10^{14}\,$W/cm$^{2}$. These calculations take into account the influence of the ionic core on the electron trajectories, which broadens the distribution of recollision angles. After summing the recolliding electrons over the polar angle, we obtain a recollision angle distribution peaked around $\pm$6$^{\circ}$ \cite{Shafir, Higuet}. In clusters, given the high charge states observed in our experiments, we expect the ionic potential to play a more important role \cite{Fukuda}. Futhermore screening effects can inhomogenize the electric field over the cluster \cite{Skopalova}. These two effects will lead to a broader recollision angle distribution. 

For understanding the origin of the apparent depolarization of harmonic emission, we performed simple simulations of the macroscopic generation process assuming an infinitely thin medium. We randomly distribute N emitters on a square grid  of 100*100$\mu$m which represents the generating medium. Each emitter located in $(x,y)$ radiates an electric field $\vec{E}_q(x,y)=I_0(x,y)^{q_{Eff}/2} e^{i\alpha_q I_0(x,y)} \vec{u}(x,y)$, where $I_0(x,y)$ is the fundamental intensity distribution, $q_{Eff}$ the effective non-linearity of the harmonic emission (typically 5), $\alpha_q$ the intensity dependence of the harmonic phase ($\alpha \approx 10\times 10^{-14}$ cm$^2$/W for the end of the plateau), and $\vec{u}(x,y)$ is unit vector with a direction randomly picked in the distribution of recolliding angles. For clusters the latter is assumed to be three times broader than the one obtained for atoms by using CTMC. We calculate the electric field resulting from the coherent sum of the contributions from the N emitters in the far field by Fourier transforming the near field profile. Two extreme cases are considered: emission from an ensemble of clusters and from an ensemble of monomers with a density of $10^{18}$ cm$^{-3}$. Assuming that only one atom out of $10^5$ emits high harmonics (due to the recombination probability) and a medium thickness of 100\,$\mu$m, we get $10^7$ atoms to distribute on our grid. For clusters, the density is much lower: assuming a medium with 80$\%$ of the atoms forming clusters of 30000 monomers, the density is $\approx 3\times 10^{13}$ cm$^{-3}$ \cite{Dorchies}, which leads to a few hundred to thousand emitters with an emission probability the same or higher than that for atoms. In the following we will consider 1000 emitters and we checked that the results were qualitatively similar using 100 to 10000 emitters. 

These simulations show that while in the case of atomic emission the harmonic far field profile is well defined in a single laser shot, the profile obtained from clusters is very inhomogeneous and fluctuates significantly from shot to shot because there are too few emitters to obtain a nice constructive interference on axis and destructive off axis (Fig. \ref{fig:polar}(b)). The polarization state from clusters (not shown) is also very inhomogeneous, showing important polarization angles and ellipticities. When averaging over 1000 shots, the spatial intensity profile becomes much smoother and the polarization angle is quasi homogeneously equal to zero. We also calculate the degree of linear polarization of the resulting light, by defining a Malus law out of the Stokes parameters and extracting its normalized oscillatory component. While the polarization appears perfectly linear in monomers, the emission from clusters appears depolarized. These results confirm that the unpolarized cutoff harmonics (Fig. \ref{fig:spec}(b)) are a fingerprint of HHG from large clusters, which have a low density in the jet and thus produce inhomogeneous harmonic emission.  

\begin{figure}
\begin{center}
\includegraphics[width=0.5\textwidth]{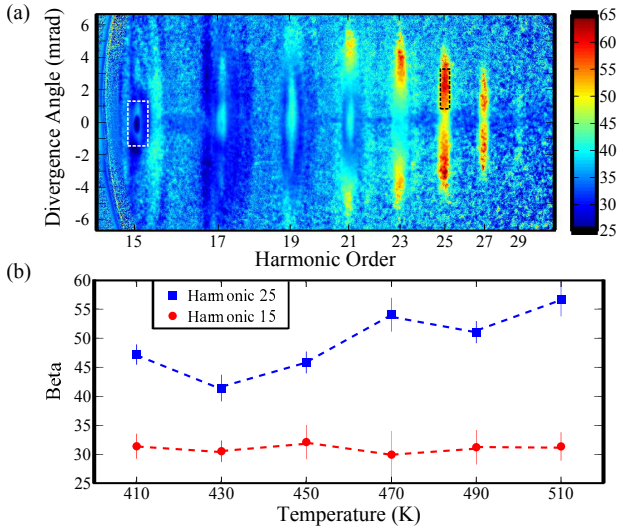}
\end{center}
\caption{(a) Harmonic decay rate $\beta$ at 510\,K. (b) Evolution of $\beta$ with temperature. The values are an average over the areas depicted above.}
\label{fig:betamap}
\end{figure}

In order to further elucidate clusters as a class of emitters, we studied the sensitivity of the high harmonic signal to the ellipticity $\epsilon$ of the generating laser field. The main polarization axis was kept vertical by employing a fixed zero-order quarter waveplate behind an adjustable zero-order half waveplate. The ellipticity was varied from -0.45 to 0.45 recording 49 points. As $\epsilon$ increases, we observe an exponential decay of the whole harmonic spectrum, which is consistent with a recollision picture of HHG. The decay rate $\beta$ of the harmonic intensity $I$ with ellipticity can be evaluated by a Gaussian fit: $I \propto exp(-\beta \epsilon^2)$. We extract this decay rate for each pixel of the detector and obtain the results shown in Fig. \ref{fig:betamap}(a). As in Fig. \ref{fig:spec}(b)) the same areas can be identified, the off-axis cutoff with $\beta>45$ and on-axis plateau with $\beta<45$. Remarkably the nozzle temperature increases the decay rate in the first case, while it hardly affects $\beta$ in the second case (Fig. \ref{fig:betamap}(b)). When extracting the slope for each pixel, a positive slope is observed only for the off-axis cutoff region.

The exact mechanism of high harmonic generation in clusters, responsible for the emission of the highest harmonics, has yet to be determined. First, let us mention that the laser intensity and electron density are too low to produce significant coherent wake emission from the plama \cite{Quere}. Second, the fast decay of the harmonic signal with ellipticity indicates that the process is recollisional and is thus different from what was recently reported in the case of bulk crystals \cite{Ghimire}, which still shows a significant signal at $\epsilon=0.5$. Third, recombination to neighbouring ions \cite{Zaretsky} can also be excluded as it would lead to lower $\beta$-values. We therefore suggest another mechanism of HHG in clusters, namely tunnel ionization from a partly delocalized electron wavefunction and recombination to this wavefunction (cluster-to-itself). Even localized electronic states in a Van der Waals clusters \cite{Feifel}, could be driven by the strong laser field from one site to another, ending up in a partially delocalized wavefunction after a few optical cycles \cite{Lein}. This process is accompanied by a significant amount of ionization of the cluster (as observed from the plasma lines) which will increase its ionization potential and consequently the higher cutoff observed for the cluster emission. Recording the harmonic signal versus $\epsilon$, one obtains the cross-correlation between the recolliding wavefunction and the ground wavefunction. In the "cluster-to-itself" picture the electrons of interest are expected to tunnel mostly from the surface of the cluster, so that the initial width of the electron wavepacket after tunneling is proportional to the cluster size. The wavepacket spreads laterally during acceleration outside the cluster -- the smaller the initial wavepacket, the stronger the spread. The decay rate, $\beta$, is determined by the width of the recolliding wavepacket and the extension of the initial wavefunction, i.e. the cluster size. For large enough clusters the latter will be dominant, so that the decay rate is expected to decrease with increasing cluster size. This is consistent with the observed temperature dependence of H27 in Fig. \ref{fig:betamap}(b). 

In conclusion, by performing a 2D spectro-spatial analysis, we are able to disentangle several contributions to the harmonic signal from a mixture of clusters and monomers. The high-energy off-axis emission shows three clear features: a higher cutoff of the harmonic emission, a higher decay rate with ellipticity when the average cluster size is reduced, and a very low degree of linear polarization. We attribute these effects to a high harmonic generation mechanism in which delocalized electrons tunnel from a cluster and recollide coherently to the whole cluster. This hypothesis needs to be further investigated using the appropriate theoretical tools. This work could be extended by performing a complete polarimetry study of the harmonic emission in elliptical laser fields, which was recently shown to be a sensitive probe of the influence of the ionic potential in HHG \cite{Shafir}. This would enable us to determine the process responsible for the generation of the lowest harmonics and possibly further differentiate the contribution from monomers and from atom-to-atom emission in clusters. The use of advanced 2D experimental characterization to disentangle contributions from different classes of emitters and different mechanisms can be applied to many situations in high harmonic spectroscopy, and should in particular be useful for other cases of HHG in inhomogeneous media, like metal plasmas \cite{Ganeev}.

We thank F. Catoire, F. Dorchies, N. Dudovich, B. Pons, F. Qu\'er\'e, H. Soifer and V. Strelkov for inspiring conversations and R. Bouillaud, C. Medina and L. Merzeau for technical assistance. We acknowledge financial support by the ANR (ANR-08-JCJC-0029 HarMoDyn) and Conseil Regional Aquitaine (20091304003 ATTOMOL and 2.1.3-09010502 COLA project).

\end{document}